\documentclass[aps,pra,twocolumn,'showpacs','showkeys',groupedaddress]{revtex4-1}

\usepackage{graphicx}
\usepackage{float}

\begin{document}


\title{All optical cooling of $^{39}$K to Bose Einstein condensation}


\author{G. Salomon}
\author{L. Fouch\'{e}}
\author{S. Lepoutre}
\author{A. Aspect}
\author{T. Bourdel}
\affiliation{Laboratoire Charles Fabry, Institut d'Optique, CNRS, Univ Paris-Sud - 2, Avenue Augustin Fresnel, 91127 Palaiseau Cedex, France}


\date{\today}

\begin{abstract}
We report the all-optical production of Bose Einstein condensates (BEC) of $^{39}$K atoms. We directly load $3 \times 10^{7}$ atoms in a large volume optical dipole trap from gray molasses on the D1 transition. We then apply a small magnetic quadrupole field to polarize the sample before transferring the atoms in a tightly confining optical trap. Evaporative cooling is finally performed close to a Feshbach resonance to enhance the scattering length. Our setup allows to cross the BEC threshold with $3 \times 10^5$ atoms every 7s. As an illustration of the interest of the tunability of the interactions we study the expansion of Bose-Einstein condensates in the 1D to 3D crossover.
\end{abstract}



\maketitle


\section{Introduction} 
As versatile, extremely well controlled, and isolated systems, ultracold atoms are widely used for quantum simulation \cite{Bloch2008}. A key feature in ultracold atoms experiments is that interactions are characterized by a unique parameter, i.e. the scattering length, which can be tuned using magnetic Feshbach resonances  \cite{Inouye1998}. In the case of fermions, changing the interaction strength has allowed the study of the BEC-BCS crossover \cite{Bourdel2004,Bartenstein2004,Zwierlein2005} as well as the thermodynamical properties of fermionic systems at unitarity \cite{Nascimbene2010,Ku2012,Tey2013}. The properties of Bose-Einstein condensates can also be greatly modified \cite{Cornish2000, Khaykovich2002, Strecker2002, Weber2003,Roati2007, Pollack2009, Haller2009}. More recently, strongly interacting Bose gases have received special interest both from experimental and theoretical points of view \cite{Kraemer2006,Zaccanti2009,Navon2011,Rem2013,Fletcher2013,Makotyn2014}. 

Due to the existence of both fermionic and bosonic isotopes as well as broad Feshbach resonances at low magnetic field, potassium is an attractive alkali \cite{D'Errico2007}. However the most common isotope - the bosonic $^{39}$K - suffers from two major limitations. The first one is a narrow hyperfine structure that hinders sub-Doppler cooling using conventional techniques. It has recently been overcome by implementing either bright molasses close to resonance \cite{Landini2011} or gray molasses on the D1 transition \cite{Salomon2013,Nath2013}. The second limitation is due to a small and negative background scattering length at zero magnetic field, which leads to a minimum of the collision cross section at relatively low energy \cite{Landini2011}. As a result, evaporative cooling in a magnetic trap has proved to be inefficient. To circumvent these problems a solution is to use other species to sympathetically  cool potassium to quantum degeneracy \cite{Roati2007,Campbell2010}. This however demands a second laser system. More recently, pioneering work combining a quadrupole trap and direct transfer to a deep optical dipole trap has led to the Bose Einstein condensation of  $^{39}$K in a single species experiment \cite{Landini2012}.

In this paper, we report the first production of a Bose Einstein condensate of $^{39}$K using an all optical method  \cite{Barrett2001, Clement2009}, i.e. by direct loading of a dipole trap from gray molasses without the aid of another atomic coolant, nor pure magnetic trapping. Optical cooling and trapping allows us to use the magnetic field to adjust the scattering length to an optimal value. This allows for a rapid production of degenerate ultracold gases with tunable interactions.

The paper is organized as follows. The direct loading of a large volume dipole trap from gray molasses is described in section II. The procedure to obtain a polarized and dense sample is then discussed in section III. The evaporation close to a Feshbach resonance is described in section IV. Finally, the tunability of the interaction is used to study the expansion of BEC in the 1D to 3D crossover in section V.  

\section{Dipole trap loading} 
The beginning of the experiment is described in \cite{Salomon2013}. $^{39}$K atoms are collected from a vapor in a two dimensional (2D) magneto-optical trap (MOT), which loads a 3D-MOT in a second ultra high vacuum chamber. The 3D-MOT saturates at $5 \times 10^{9}$ atoms after a $2.5$\,s loading stage. The cloud is further compressed in a hybrid compressed MOT (CMOT) configuration combining light on the D1 and the D2 transitions. A gray molasses working on the D1 transition  \cite{Boiron1996,Grier2013,Salomon2013} then cools the gas down to $T=6\,\mu$K.
 \begin{table*}[ht]
\centering 
 \begin{small}
\begin{tabular}{| c | c | c | c | c | c |} 
  \hline
 
Experimental steps & N &  T($\mu$K) &  n (cm$^{-3}$) & $\gamma$ (G.cm$^{-1}$) & $B_{0}$(G) \\ [1ex] 
\hline 
D1 gray molasses  & $2\times10^{9}$  & 6 &  $1 \times 10^{11}$  & 0 & 0 \\ 
Polarized sample in large trap & $8\times10^{6}$& 9 & $3.2\times 10^{11}$ & 14 & 0\\
Confining trap at the beginning of the evaporation & $4\times 10^{6}$ & 220 & $7.5 \times 10^{13}$ & 0 & 550.5 \\[1ex]
\hline 
\end{tabular}
\end{small}
\caption{Experimental parameters during the steps preceding evaporation.  
We report the atom number $N$, the temperature $T$, the density $n$, the magnetic field gradient in the strong direction $\gamma$, and the Feshbach field $B_{0}$.}
\label{parameters} 
\end{table*}

The atoms are loaded in a large dipole trap derived from a $30\,$W multimode IPG fiber laser ELR-30-1550-LP operating at $1550\,$nm. The $21\,$W beam is focused down to a  waist radius at $1/e^2$ of $\omega_1=150\,\mu$m and shined on the atoms at the beginning of the molasses phase  \cite{Clement2009, Jacob2011, Burchianti2014}. It creates a $U_1 = k_B \times 40\,\mu$K deep optical potential ($k_B$ is the Boltzmann constant). The highest trapped atom number of $3 \times 10^7$ is obtained for the laser parameters leading to the coldest gray molasses. This is expected as the induced light shifts of the order of a few megahertz are small compared to the laser detunings and moreover do not affect the Raman condition between the hyperfine ground states in the gray molasses \cite{Salomon2013}. 

As compared to a sudden loading at the end of the molasses phase, the superposition of the optical trap with the gray molasses leads to a doubling of the number of captured atoms. We cannot collect more atoms using longer superposition times than $\sim 10\,$ms, an effect which we interpret as a density dependent efficiency of the gray molasses limited by photon reabsorption (above $\sim 2 \times 10^{11}\,$cm$^{-3}$) \footnote{Our interpretation is in agreement with the observation of strong losses when we try to recool the atoms using gray molasses after compression in a deep optical trap}. Our transferred atom number of $3 \times 10^7$ roughly matches an estimation given by the number of atoms being at a position where their kinetic energy does not exceed the local trap depth \footnote{We can for example evaluate the number of atoms contained in the volume for which the trap depth exceeds the thermal energy $k_{B}T$. It is a cylinder of radius $\rho=\omega_1 \sqrt{\text{ln}(U/k_{B}T)/2}$. A more complex integration of the cloud in phase space gives a similar result.}. This is consistent with our optimization of the captured atom number by increasing the trap waist up to 150\,$\mu$m. For even larger waists, gravity plays in important role and hinders an efficient trapping.

\section{Sample polarization and compression} 
During the gray molasses the atoms are distributed among the different $F=1$ and $F=2$ ground states.
In the very last 0.5\,ms of the gray molasses, the repumping light is switched off in order to optically pump the atoms in the $F=1$ sublevels, forming a stable mixture. Due to the large Rayleigh length of the optical dipole trap of 4.5\,cm, the atoms are not confined along the beam direction. The initial (rms) size of the cloud is in this direction is 1\,mm, {\it i.e.} the size of the gray molasses. In order to perform an efficient evaporation at a high collision rate, we thus need to confine and compress the cloud to a higher density. This can for example be done by directly adding another crossing optical dipole trap \cite{Clement2009}. In the case of potassium, this is not sufficient because of the scattering length a$_{bg} \approx -30 \times $a$_{0}$ where a$_{0}$ is the Bohr radius. Indeed this low and negative value implies the existence of a Ramsauer collision rate minimum at a relatively low collision energy of $k_B \times 400 \,\mu$K \cite{Landini2012}. When compressing the gas, the temperature increases towards this energy and prevents the evaporation from being efficient. 

To circumvent this problem, we take advantage of $^{39}$K Feshbach resonances \cite{D'Errico2007} to tune the scattering length. We thus need to polarize the gas in a single spin state because there is no magnetic field region where the scattering lengths for the different $F=1$ sublevels are simultaneously favorable to initiate an efficient evaporation. Experimentally the optical pumping efficiency is limited to 60$\%$ due to a too high density at the end of the gray molasses. Moreover optical pumping leads to residual heating and to a factor 4 decrease in the atom number captured in the optical trap \footnote{We have also tried optical pumping in the optical trap after loading. This turns out to be inefficient due to a higher atomic density.}. 

\begin{figure}[h]
 \centering
 \includegraphics[width=0.45\textwidth]{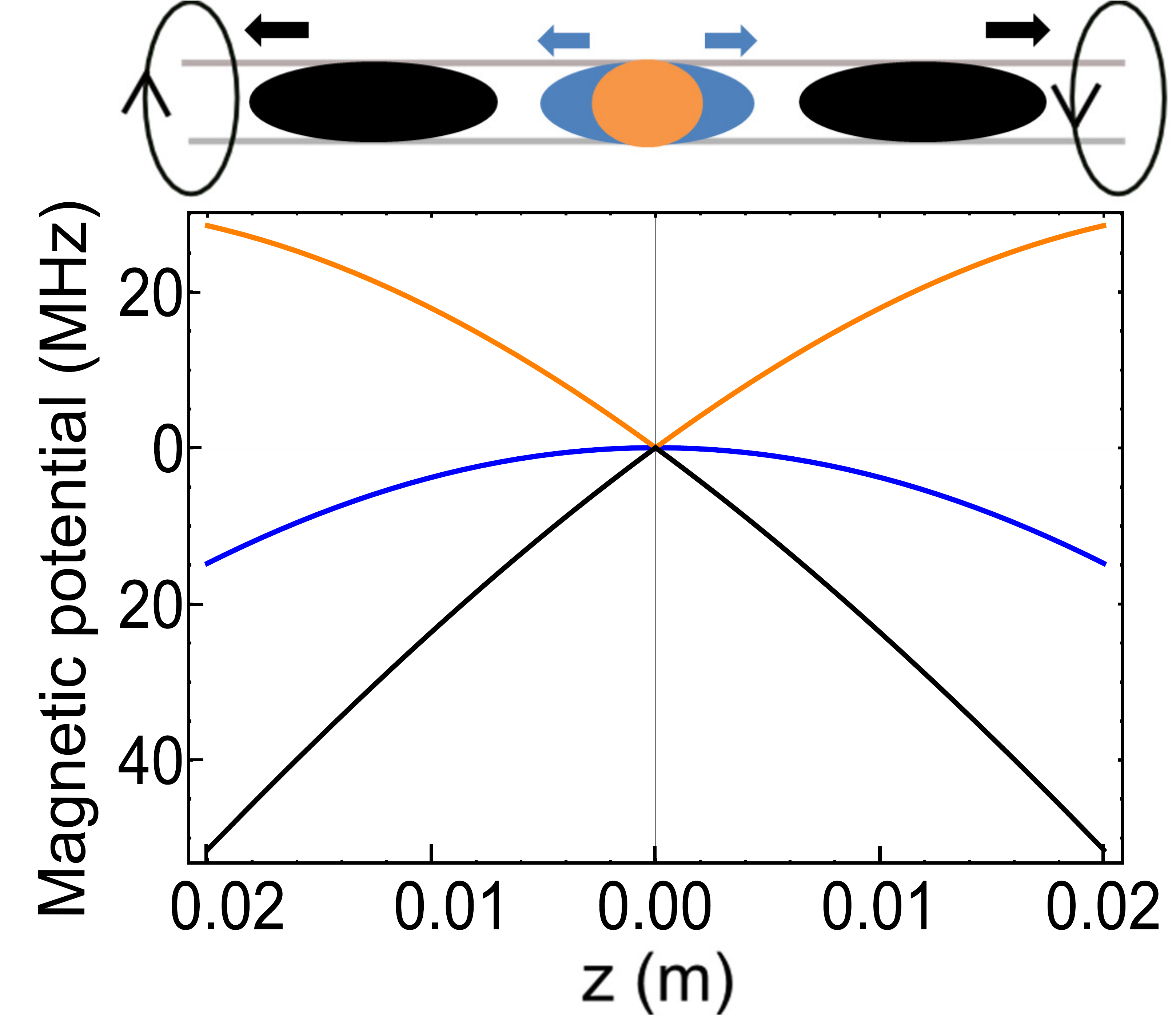}
 \caption{Sample polarization procedure in a quadrupole field. Each Zeeman sublevel (represented with different colors) experiences a different longitudinal potential. It is a confining potential only for the $m_F=-1$ (in light orange). The two others are repulsive and given by the second order Zeeman shift for the $m_F=0$ (in blue), and by the first order Zeeman shift for the $m_F=+1$ (in black). \label{trimf}}
 \end{figure}

We thus adopt another strategy that is filtering out unwanted atomic levels. In practice, we apply, in addition to the optical trap, a small magnetic quadrupole field of $14\,$G.cm$^{-1}$ using the MOT coils.  In this configuration, only the atoms in the $|F=1,m_F=-1\rangle$ state experience magnetic confinement along the beam direction, while the others are lost due to a repulsive magnetic potential (see fig.\,\ref{trimf}). This configuration allows us, after 500\,ms, to keep one third of the initially trapped atoms in a pure spin state. 

A second $13$W dipole trap beam with a $21\,\mu$m waist radius is then crossed at the centre of the large trap with an angle of $56$ degrees \cite{Clement2009}. The atoms from the first beam are transferred into this second highly confining trap on a time scale of 1\,s. After $2$s the magnetic quadrupole is switched off and we are left with $N=4 \times 10^{6}$ atoms in the $|F=1,m_F=-1\rangle$ state at $220\,\mu$K in this tightly confining optical trap (second beam). Since the temperature is then higher than the trap depth of the large trap, the latter has no important effect in the cloud trapping. At this stage, the trapping frequencies can thus be approximated by the ones of the confining trap, $\omega_{\parallel} / 2\pi= 145(7)\,$Hz in the longitudinal direction and $\omega_\perp / 2\pi= 8.7(0.5)\,$kHz in the radial one. The resulting phase space density $D=N\omega_\perp^2 \omega_{\parallel} (\hbar/k_B T)^3$ is $4.5\times10^{-4}$. Note that, even in this highly confining dipole trap at 1550\,nm, we do not observe any light induced losses. This is in contrast to the situation observed with a multimode laser at $1070\,$nm  \cite{Landini2012}.

\section{Evaporative cooling in the optical trap} 
\begin{figure}[h]
\centering
\includegraphics[width=0.52\textwidth]{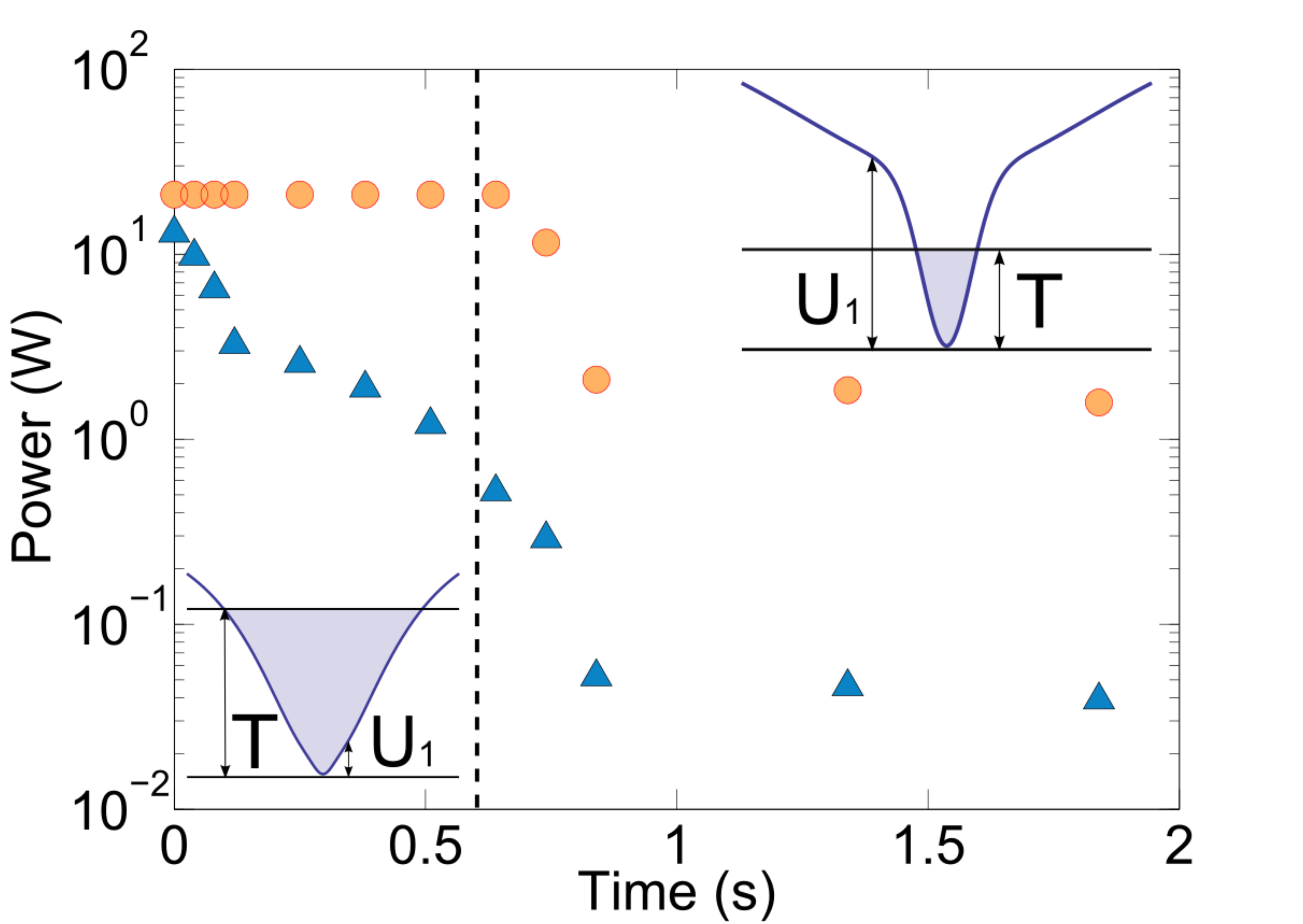}
\caption{Power reduction during the evaporation process of the large beam (orange circles) and of the tightly confining one (blue triangles). During the first part of the evaporation the atoms mainly see the second (confining) trap. Indeed $k_BT \gtrsim U_1$ where $U_1$ is the depth of the large beam. The potential is changed by reducing the power of the confining beam and the atoms enter the crossed region when $k_BT \lesssim U_1$. The large beam then determines the longitudinal frequency. The crossover between the two trapping regimes is shown by a dashed line. The powers of both beams are reduced together during the remaining part of the evaporation. \label{ramps}}
\end{figure}

In order to perform efficient evaporative cooling to quantum degeneracy, 
we first adjust the magnetic field to $550\,$G in order to tune the scattering length to $a=130\times$a$_{0}$ \cite{D'Errico2007}. It increases the collision rate to 15\,ms$^{-1}$. The forced evaporation proceeds as follows (see fig.\,\ref{ramps}). In the first part of the evaporation the power of the tightly confining trap is decreased by a factor $\sim 25$ until the trap depths of the two beams are of the same order. In the second part of the evaporation the power of both beams is decreased simultaneously.  The condensation transition is crossed with $3\times10^5$ atoms after 2\,s of evaporation. 

\begin{figure}[ht]
 \centering
 \includegraphics[width=0.5\textwidth]{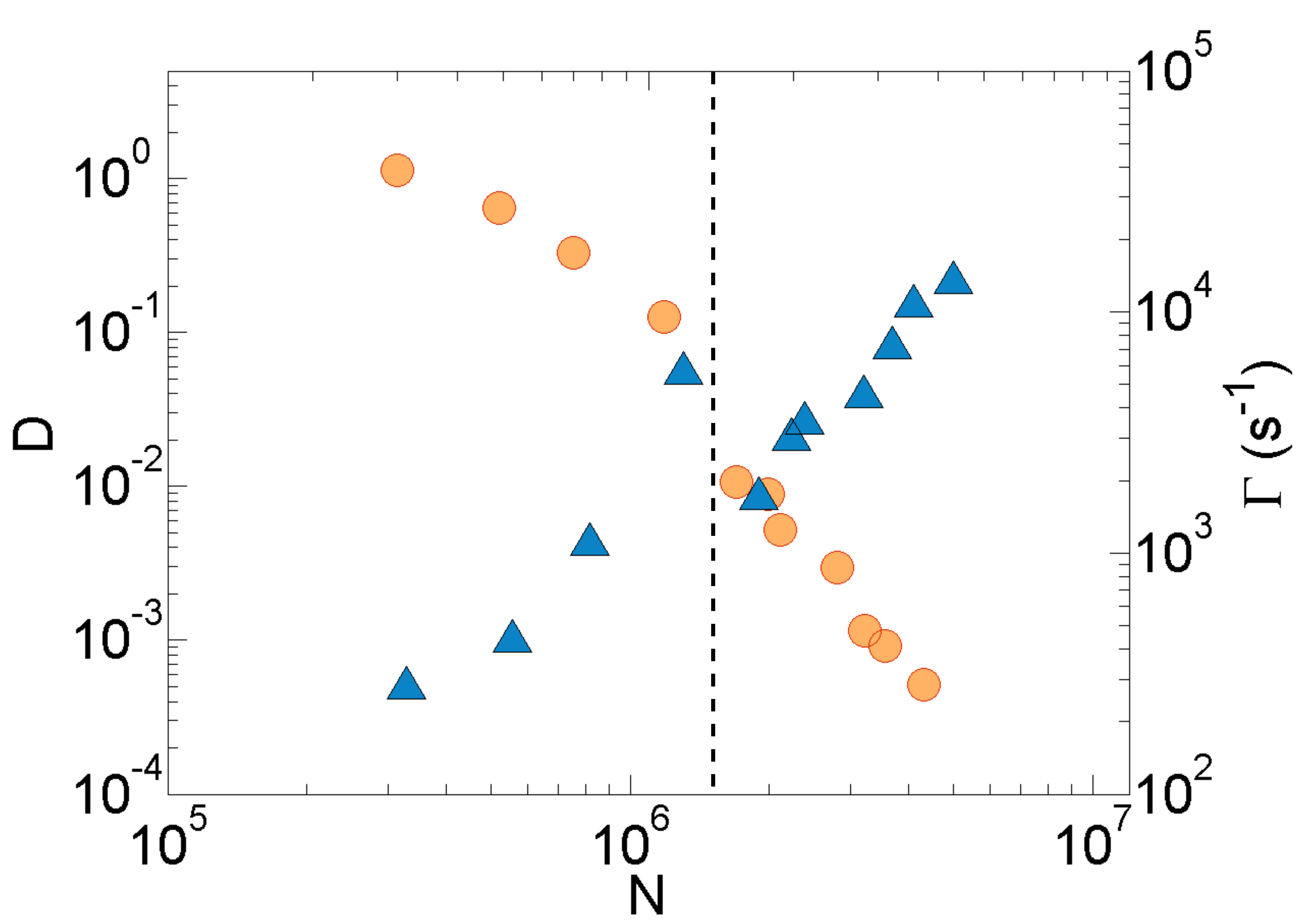}
 \caption{Phase space density (orange circles) and two body collision rate $\Gamma$ (blue triangles) during the evaporation process as functions of the atom number. The corresponding average efficiency is $\varrho\sim 3$. The dashed line represents the stage at which the atoms enter the crossed region of the traps as explained in fig.\,\ref{ramps} \label{PSD}}
 \end{figure}

In order to evaluate the efficiency of our evaporation, we plot in figure \ref{PSD} the phase space density $D$ and the collision rate $\Gamma=\sqrt{2/\pi} \, n_0 \,(8 \pi a^2)\,\sqrt{k_BT/m}$, where $n_0=N\omega_\perp^2 \omega_{\parallel} (m/2\pi k_B T)^{3/2}$ is the central density, and $m$ the atomic mass \cite{Dalibard2006}. The temperature and the number of atoms in the gas are measured through the size of the cloud after a time of flight of 20\,ms. The trap frequencies are deduced from the independently measured properties of the two dipole trap beams. In our analysis, we assume that the longitudinal trap frequency is either given by the longitudinal frequency of the tightly confining beam or given by the radial confinement of the large beam depending if the atoms predominantly occupies the cross region (see Fig.\,\ref{PSD}) . 

To quantify the efficiency of the evaporation process we use the derivative $\varrho=-\mathrm{d log}D / \mathrm{d log}N$. We measure an average efficiency $\varrho\sim 3$ (see fig.\,\ref{PSD}). The collision rate always remains larger than some hundreds per second which is high enough for an efficient evaporation. The increase of the collision rate around $10^6$ atoms corresponds to the transfer of the atoms in the crossed region \footnote{In contrast to \cite{Clement2009}, we do not use an off centered configuration. It was not necessary to achieve an efficient evaporation.}. It appears as a sudden jump because of our approximation of an abrupt increase in the longitudinal frequency of the trap.

Our evaporation in an optical trap is both efficient and fast. It takes 2\,s to reach the condensation threshold while losing a factor $\sim$10 in atom number. The chosen scattering length is optimal and changing it to a higher or lower value during the evaporation does not improve our efficiency. We could also evaporate with similar efficiency when using the same scattering length close to the two other low field Feshbach resonances at 33\,G and 162\,G.

\section{Condensate with tunable interactions}

In this section, we use our ability to tune the scattering length to study Bose-Einstein condensates in the 1D to 3D crossover. More precisely, we study the expansion of condensates released from an elongated trap at different magnetic fields. At the end of the evaporation producing Bose-Einstein condensates of $N_c \sim 2\times 10^4$ atoms, the trapping frequencies are $ \omega_{\parallel} / 2\pi= 24(1)\,$Hz in the longitudinal direction and $\omega_\perp / 2\pi= 305(10)\,$Hz in the radial one. The magnetic field is first linearly ramped from 550\,G to values between 400\,G and 555\,G, which corresponds to scattering lengths between $-30\,$a$_0$ and $300\,$a$_0$. After an additional 200\,ms equilibration time, we then switch off the optical trap and let the gas freely expands. After 10\,ms the interaction energy is already converted into kinetic energy and we can switch off the magnetic field without modifying the expansion \footnote{If we switch the trap and the magnetic field simultaneously, we observe losses associated with the crossing of the low field Feshbach resonances \cite{Stenger1999,Cornish2000}.}. After $t=34.5\,$ms of expansion, the condensates are finally imaged at zero field by fluorescence using the MOT beams at resonance during 50\,$\mu$s. We observe the onset of three body losses when the scattering length is set to values higher than $300\,$a$_0$. At magnetic fields lower than 505\,G, $i.e.$ for $a<0$, the collapse of the Bose-Einstein condensate is observed through its disappearance \cite{Sackett1999}. 

\begin{figure}[ht]
 \centering
 \includegraphics[width=0.52\textwidth]{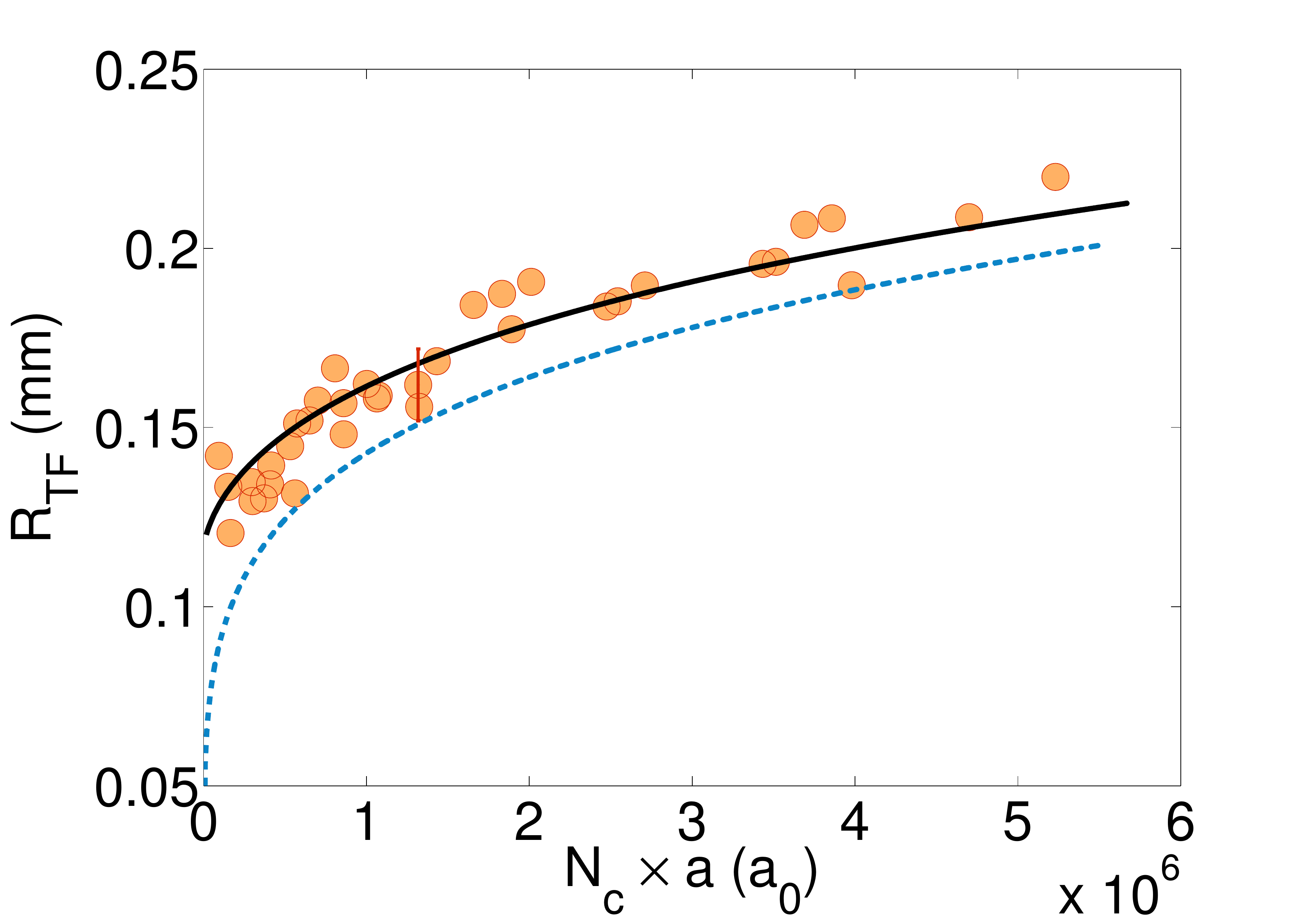}
 \caption{Experimental radial Thomas Fermi size after 34.5 ms of expansion as a function of $N_c a$, where $N_c$ is the number of condensed atoms. The theoretical curve (continuous blue line) corresponds to the size given by an approximated theory in the 1D to 3D crossover. The blue dashed line corresponds to the Thomas-Fermi model in 3D which does not match our data. \label{expansion}}
 \end{figure}

For $10<a<300\,$a$_0$, we study the radial size of the condensate after expansion. Figure \ref{expansion} shows this size as a function of the parameter $N_c a$ which characterizes the strength of the interactions. We observe that due to the increase of the interaction energy, the size after expansion grows when $N_c a$ increases. In practice, we have chosen to fit the shape of the condensate with a Thomas-Fermi profile  \cite{Castin1996, Kagan1997, Dalfovo1999}. 

Because $\mu$ is not much larger than $\hbar  \omega_\perp$, we are in the 1D-3D crossover and the Thomas-Fermi theory is not applicable.  We have thus developed a model to adjust our data. It is based on an accurate interpolated value of the local chemical potential in the 1D-3D crossover in a radially trapped cloud $\mu=\hbar \omega_\perp \sqrt{1+4n_{1D}a}$, where $n_{1D}$ is the 1D density along the longitudinal direction \cite{Gerbier2004}. This formula interpolates between $\mu=\hbar \omega_\perp$, in the 1D regime, and $\mu=\hbar \sqrt{4n_{1D}a}$, which corresponds to the 3D Thomas-Fermi regime. In a cylindrical section of length $l$, the number of atoms is $n_{1D}l$ and the energy is
\begin{equation}
E=l \int_0^{n_{1D}} \mu(n) \textrm{d}n=l \hbar \omega_\perp (1+4n_{1D}a)^{3/2} /6a.
\end{equation}
Thanks to the virial theorem in 2D \cite{Werner2008}, half of this energy is potential and the other half is the sum of the kinetic and interaction energies, which is the release energy in expansion $E_{rel}$. Using the local density approximation in the longitudinal direction, we can then calculate the radial release energy per particle.  In order to compare with the data, we finally compute the corresponding Thomas-Fermi radius after expansion $R=\sqrt{7E_{rel} / m} \times t$. The theoretical curve matches well our data. Note that the latter are fitted with a single free parameter that is the atom number calibration or in other words the detection efficiency. The obtained scaling factor is consistent with the atom number calibration obtained via the study of the condensed fraction as a function of $T/T_c^0$ around the transition. We estimate a $30\%$ uncertainty on the measured number of atoms.

\section{Conclusion}

We have demonstrated a new route to produce $^{39}$K BECs in a single species experiment. It is based on an all optical method that allows for a fast production of degenerate quantum gases. The atoms are directly loaded in a dipole trap from D1 gray molasses. After filtering out unwanted Zeeman substates evaporation is performed close to a Feshbach resonance with a high efficiency. It results in producing BECs with tunable interactions every $7\,$s in our setup. As an example we have finally studied the expansion of a BEC in the 1D to 3D crossover while tuning the scattering length finding good agreement with theory.  In future works, the atom number in the BEC can be increased by using more power in the dipole trap thus allowing the use of larger waists. Our method will facilitate future studies of degenerate Bose gases with tunable interactions and will find applications in both interferometry as well as quantum simulation.

\begin{acknowledgments}
We acknowledge P. Bouyer and P. Wang for important experimental contributions, and F. Moron and A. Villing for technical assistance. This research was supported by CNRS, Minist\`ere de l'Enseignement Sup\'erieur et de la Recherche, Direction G\'en\'erale de l'Armement, ANR-12-BS04-0022-01, RTRA: Triangle de la physique, iSense, ERC senior grant Quantatop, DIM Nano'K from region Ile-de-France. LCFIO is member of IFRAF.
\end{acknowledgments}

\bibliography{biblio}

\end{document}